\documentclass[aps,prl,twocolumn,superscriptaddress,nobibnotes]{revtex4-2}
\usepackage{amsmath,amssymb}
\usepackage{hyperref} 
\usepackage{graphicx}
\usepackage{float}
\usepackage{braket}
\usepackage{xcolor}
\usepackage{placeins}

\begin{document}
	
	\title{The unique control features of topological stochastic and quantum systems}
	
	\author{Ziyin Xiong}
    \affiliation{Department of Physics, Rice University, Houston, TX 77005, USA}
    \affiliation{Department of Physics, Massachusetts Institute of Technology, Cambridge, MA 02139, USA}
    \affiliation{Center for Theoretical Biological Physics (CTBP), Rice University, Houston, TX 77005, USA}
	
	\author{Aleksandra Nelson}
	\affiliation{Center for Theoretical Biological Physics (CTBP), Rice University, Houston, TX 77005, USA}
	
	\author{Evelyn Tang}
    \email{e.tang@rice.edu}
	\affiliation{Department of Physics, Rice University, Houston, TX 77005, USA}
    \affiliation{Center for Theoretical Biological Physics (CTBP), Rice University, Houston, TX 77005, USA}

	\begin{abstract}
		Topological phases support edge states that can be robust to material deformations and other perturbations. While well-studied in quantum systems, topological phases have also been observed in stochastic and biochemical systems, yet it remains unclear which of their properties remain similar or different from those in quantum systems. In this paper, we derive analytical expressions for the spectral properties of simple quantum and stochastic models on the same lattice to rigorously characterize these complex systems. Intriguingly, we find that non-reciprocity moves states away from the steady-state in stochastic systems while clustering states at zero-energy in quantum systems. In contrast, making the system more topological does the opposite: it clusters more states around the steady-state in stochastic systems but moves states away from the zero-energy state in quantum systems. These results provide control parameters for selection and modulation of different purposes while quantifying the size of gap which protects the longest-lived states. Lastly, we discover a mode unique to stochastic systems that we dub the topologically emerging state, which persists across different models and dimensions, including in the presence of non-equilibrium currents.

	\end{abstract}
	
	\maketitle
	
	\section{Introduction}

Topological phases of matter provide a powerful framework for understanding robust phenomena in quantum systems. In particular, topological insulators and related models can support boundary-localized edge states that persist under disorder and geometric deformations as long as a protecting spectral gap remains open \cite{mahault2022topo,sheng2005spin,liu2017zero,zhu2021delocalization}. 
These properties are typically diagnosed through spectral features of the Hamiltonian, including band gaps, winding numbers, and the presence of protected boundary modes \cite{jorgens2006spectral,tran2009accurate}.

Recently, analogous topological phenomena have been explored beyond equilibrium quantum systems, including in non--hermitian and stochastic settings \cite{okuma2023non,ding2022non,borgnia2020non}. 
Non--hermitian Hamiltonians naturally arise in open or driven systems with gain and loss, and can exhibit unique features such as the non--hermitian skin effect, in which an extensive number of eigenstates become boundary-localized under non-reciprocal hopping \cite{kawabata2019symmetry,lee2016anomalous,yao2018edge,song2019nonhermitian,song2019realspace,yin2018geometrical}. 
Meanwhile, stochastic and biochemical networks are intrinsically open and dissipative: energy consumption and thermodynamic forces can break detailed balance, generating non-reciprocal probability currents in state space \cite{schnakenberg1976,amir2016nonhermitian,sawada2024topology,Nelson2024}. 

Beyond identifying topological signatures, a central challenge is understanding how non-equilibrium systems can be \emph{controlled}: which parameters govern the existence, robustness, and lifetime of long-lived boundary modes. Because eigenmodes governing relaxation toward the steady state shape response to perturbations, recent work establishing universal thermodynamic bounds on non-equilibrium fluctuations in stochastic systems further motivates understanding how spectral structure can be tuned \cite{gingrich2016dissipation,owen2020universal,horowitz2020thermodynamic}.

A natural starting point for the comparison between quantum and stochastic descriptions is the spectral structure of the quantum and the stochastic operators. The close relationship between stochastic generators and quantum Hamiltonians has been explored in stochastic mechanics, where the Schrödinger equation emerges from Brownian motion dynamics \cite{Nelson1966}. Further, for models defined on the same graph, the quantum operator is naturally expressed in terms of an adjacency matrix, while the stochastic generator is closely related to a graph Laplacian \cite{newman2018networks,wong2016laplacian}. Both the adjacency matrix and the Laplacian are well-studied operators in network control theory, e.g., for linear control with the former and for consensus or synchronization with the latter \cite{mesbahi2010graph,liu2016control}. These varying contexts motivate a careful study of both operators and the differences between them. 

Our goal is to develop a systematic understanding of how non--hermitian and stochastic topological systems can be controlled. Although topological phenomena have been observed in stochastic and biochemical systems \cite{marsland2019tur,murugan2017topo,tang2021topo,zheng2024topomech}, it remains unclear how key spectral signatures compare across quantum and stochastic settings. In this work, we identify spectral features that persist across dimensions and non-equilibrium steady states. We show that non-reciprocity and topology act as distinct control parameters that drive opposite spectral responses in quantum and stochastic systems. Our results show that minimal non--hermitian stochastic lattices possess rich, designable spectral structure and provide analytical control parameters governing robustness and relaxation in stochastic transport, biochemical regulation, and active systems.

\begin{figure*}[t]
	\centering
	\includegraphics{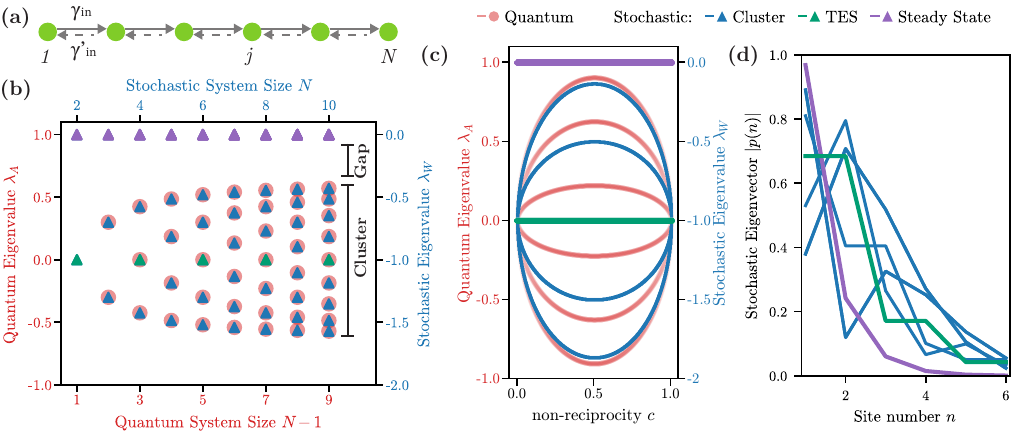}
	\caption{\textbf{Properties of 1D Hatano-Nelson non--hermitian quantum and stochastic lattices.}
		(a) A uniform lattice with open boundary conditions and non-reciprocal transition rates, parameterized by the forward rate $\gamma_{\mathrm{in}}$ and the backward rate $\gamma_{\mathrm{in}}^{\prime}$. The strength of non-reciprocity is characterized by the ratio $c = \frac{\gamma_{\mathrm{in}}}{\gamma_{\mathrm{in}} + \gamma'_{\mathrm{in}}}$. Here $N$ is the total number of sites in the chain and $j$ labels a lattice site.
        (b) The stochastic cluster for system size $N$ overlaps with the quantum cluster of size $N{-}1$. In the stochastic system, as the system size $N$ increases toward the thermodynamic limit, the spectral gap between the cluster and the steady state saturates to a constant value. Red circles denote the spectrum of $\mathcal{A}$. Triangles denote the spectrum of $\mathcal{W}$, with blue corresponding to the spectral cluster, purple to the steady state, and green to the topologically emerging state. The last state is explored in Sec.~\ref{sec:pushed}. The stochastic gap is defined as the distance between the steady-state eigenvalue and the second-largest eigenvalue. Here $c = 0.8$.
        (c) We show the stochastic and quantum spectrum as a function of $c$. As the non-reciprocity parameter $c$ increases, both quantum and stochastic clusters collapse to a single point, while the stochastic steady-state eigenvalue remains fixed. Here $N = 6$.
		(d) Stochastic eigenvectors exhibit an exponential envelope with sinusoidal modulation. Stochastic eigenvector profiles at $c = 0.8$ for $N = 6$. 
	}
	\label{fig:lattice}
\end{figure*}

\section{Formal background}
Non-equilibrium stochastic systems are often described by Markov generators governing probability flows and entropy production \cite{Nardini2017}. Stochastic systems evolve according to the master equation \cite{schnakenberg1976},
\begin{equation}
\frac{d\mathbf{P}}{dt}=\mathcal{W}\mathbf{P},
\end{equation}
where $\mathbf{P}$ is the probability vector and $\mathcal{W}$ is the transition-rate matrix. 
The eigenvalues of $\mathcal{W}$ determine relaxation timescales: eigenmodes with larger negative real parts decay rapidly, while modes closer to zero correspond to slower processes. 
Probability conservation enforces that $\mathcal{W}$ has a zero eigenvalue corresponding to the steady state \cite{su1979solitons}. 

This structure parallels the role of the Hamiltonian $H$  in quantum mechanics \cite{Nelson2024, tang2021topo},
\begin{equation}
i\frac{d\psi}{dt}=H\psi,
\end{equation}
where the spectrum of $H$ determines coherent dynamics and energy bands. Hence, the quantum and stochastic descriptions can be formally mapped through \(H = i\mathcal{W}\), so that both operators share the same eigenvectors \cite{tang2021topo,sawada2024topology,Nelson2024}. 

Despite this formal correspondence, the two spectra can be quite different even when describing the same network \cite{Nelson2024}. This is largely due to the constraint of probability conservation mentioned above for stochastic systems, that their quantum counterparts do not have. For instance, tight-binding Hamiltonians consist of off-diagonal terms (hopping between sites), which can be described by an adjacency matrix $\mathcal{A}$. In contrast, on the same network $\mathcal{W}$ would take the form of $\mathcal{W}=\mathcal{A}-\mathcal{D}$, where  $\mathcal{D}$ is a diagonal matrix that enforces probability conservation and $\mathcal{D}_{ij}=\delta_{ij}\sum_k\mathcal{A}_{ki}$ \cite{schnakenberg1976}. 

Further, in the stochastic system, the state of interest is the steady state associated with the zero eigenvalue of $\mathcal{W}$ \cite{sawada2024topology}. Concurrently, in the quantum system we analyze the zero-energy eigenstate of $\mathcal{A}$, which plays the role of the topological edge mode of the SSH chain \cite{Nelson2024}. These differences strongly shape their eigenvalue structure and long-time behavior. In what follows, we perform a comparison between the stochastic and quantum systems through the stochastic operator $\mathcal{W}$ and the quantum operator $\mathcal{A}$ by examining the physically relevant eigenstates in each framework.

\section{1D Non--Hermitian Hatano-Nelson Chain}
\label{sec:1D}
We start from a simple one-dimensional lattice to establish a baseline for comparison with more complex systems. Hence, we consider a non--hermitian tight-binding Hamiltonian (see Fig.~\ref{fig:lattice}(a))
\begin{equation}
\mathcal{A}_{\mathrm{uniform}}
= \sum_{j=1}^{N}
\gamma_{\mathrm{in}}\,|j+1\rangle\langle j|
+ \gamma_{\mathrm{in}}^{\prime}\,|j\rangle\langle j+1|,
\end{equation}
where $\gamma_{\mathrm{in}}$ and $\gamma_{\mathrm{in}}^{\prime}$ denote forward and backward hopping amplitudes, respectively. $j$ labels the unit cell, each containing a single site. When $\gamma_{\mathrm{in}}\neq\gamma_{\mathrm{in}}^{\prime}$, the Hamiltonian is non--hermitian, corresponding to non-reciprocal rates under open boundary conditions. This model is commonly known as the Hatano--Nelson chain and represents one of the simplest non--hermitian systems exhibiting nontrivial spectral and localization properties \cite{hatano1996localization,bergholtz2021exceptional}.

We first study finite system sizes to develop analytical results. Since this system is described by a Toeplitz matrix, the solution is that of Chebyshev polynomials (see details in Appendix.~\ref{sec:1Danaspec}). We thus obtain the quantum spectrum as
\begin{align}
\lambda_\mathcal{A}
&= -2\sqrt{\gamma_{\mathrm{in}}\gamma_{\mathrm{in}}^{\prime}}
\cos\!\left(\frac{k\pi}{N+1}\right) ,
&\quad k=1,2,\dots,N.
\end{align}

Then, the stochastic spectrum (transient states only) is
\begin{equation}
\begin{aligned}
\lambda_\mathcal{W}
&= -(\gamma_{\mathrm{in}}+\gamma_{\mathrm{in}}^{\prime})
+ 2\sqrt{\gamma_{\mathrm{in}}\gamma_{\mathrm{in}}^{\prime}}
\cos\!\left(\frac{k\pi}{N}\right), \\
&\qquad k=1,2,\dots,N-1.
\end{aligned}
\end{equation}

In addition, probability conservation guarantees a steady-state eigenvalue
\begin{equation}
\lambda_{\mathrm{ss}} = 0,
\end{equation}
which is separated from the rest of the spectrum.

\subsection{Spectral correspondence between quantum and stochastic spectra of neighboring sizes}
We are interested in how spectral structure and spatial localization behavior differ between quantum and stochastic models. First, we can see above using a uniform spectral shift by $-(\gamma_{\mathrm{in}}+\gamma_{\mathrm{in}}^{\prime})$, that the stochastic cluster of a system with $N$ sites overlaps with the quantum cluster of a system with $N-1$ sites, yielding an interesting dimensional relation between the two systems. We also show this in Fig.~\ref{fig:lattice}(b).

We can define the \emph{cluster size} as the difference between the largest and smallest eigenvalues for the quantum system, and between the largest non-zero eigenvalue and smallest eigenvalues for the stochastic system. For the quantum system, the spectrum forms a single continuous cluster of eigenvalues (see Fig.~\ref{fig:lattice}(b)). In contrast, the stochastic spectrum consists of a cluster separated from the steady state by a finite gap.

As the system size increases, eigenvalues in both clusters become denser, and the cluster widths asymptotically converge to the same value.

\subsection{Non-reciprocity expands the stochastic gap while clustering quantum states at zero-energy}
The degree of non-reciprocity is controlled by the ratio
$\frac{\gamma_{\mathrm{in}}}{\gamma_{\mathrm{in}}^{\prime}}$,
which can be interpreted as an effective thermodynamic driving force in biochemical networks \cite{tang2021topo,dinelli2023non,ouazan2023self}. For convenience, we adopt an alternative parameter chirality $c$ (or non-reciprocity) similar to previous papers \cite{tang2021topo,Nelson2024} and defined as:
\begin{equation}
\frac{\gamma_{\mathrm{in}}}{\gamma_{\mathrm{in}}^{\prime}}
= \frac{c}{1-c}.
\end{equation}

The system is reciprocal at $c=0.5$, where forward and backward rates are equal, whereas deviations from $c=0.5$ break chiral symmetry and introduce directional bias in the transitions, with increasing asymmetry as $c \to 0$ or $c \to 1$.

As non-reciprocity increases, both quantum and stochastic spectral clusters collapse toward a single point (see Fig.~\ref{fig:lattice}(c)). However, while the quantum eigenvalues converge toward a single point at the edge state, in the stochastic system the steady state remains fixed and is separated from the point-like cluster by a finite spectral gap. As a result, the stochastic spectral gap increases with non-reciprocity, separating the steady state from the rest of the modes.

\subsection{Eigenvector profile and spatial localization}
To quantify spatial localization, we analytically solve for the eigenvectors for both $\mathcal{A}$ and $\mathcal{W}$. Each lattice site is labeled by \( n = 1,\ldots,N \).

The quantum eigenvectors take the form:
\begin{equation}
v_n^{(k)}
=
\left(\frac{\gamma_{\mathrm{in}}'}{\gamma_{\mathrm{in}}}\right)^{\frac{n-1}{2}}
\sin\!\left(\frac{nk\pi}{N+1}\right),
\qquad n=1,\ldots,N.
\end{equation}
where $k$ labels the eigenmode and $n$ the lattice site.

The eigenvectors in the stochastic cluster are
\begin{equation}
\begin{aligned}
p^{(k)}_{\mathrm{cluster},n}
&=\Big(\tfrac{\gamma_{\mathrm{in}}}{\gamma'_{\mathrm{in}}}\Big)^{\!\frac{n-1}{2}}
\!\left[
\sin\!\big(\tfrac{nk\pi}{N}\big)
-\sqrt{\tfrac{\gamma'_{\mathrm{in}}}{\gamma_{\mathrm{in}}}}\,
\sin\!\big(\tfrac{(n-1)k\pi}{N}\big)
\right],\\
&\quad k=1,\ldots,N-1.
\end{aligned}
\label{eq:cluster_eigenvector}
\end{equation}

The analytical solution for the steady-state eigenvector has been studied in previous literature and is shown to follow an exponential profile set by the ratio of the forward and backward rates\cite{Nelson2024}.

\begin{figure*}[th!]
	\centering
	\includegraphics{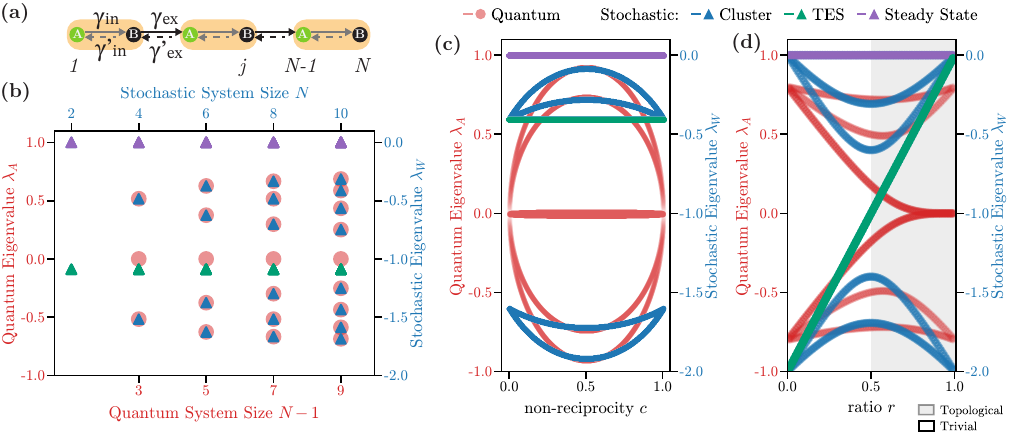}
	\caption{\textbf{Properties of 1D SSH non--hermitian quantum and stochastic lattices.}
        (a) A 1D lattice with non-reciprocal transition rates, parameterized by the alternating forward internal rate $\gamma_{\mathrm{in}}$, forward external rate $\gamma_{\mathrm{ex}}$, backward internal rate $\gamma_{\mathrm{in}}^{\prime}$, and backward external rate $\gamma_{\mathrm{ex}}^{\prime}$. The strength of non-reciprocity is characterized by the ratio $\frac{\gamma_{\mathrm{in}}}{\gamma_{\mathrm{in}}^{\prime}}
= \frac{c}{1-c}$. The yellow shading indicates one unit cell.
        (b) The stochastic cluster of system size $N$ (integer number of unit cells) overlaps with the quantum cluster of size $N{-}1$ (non-integer number of unit cells). Red circles show the spectrum of $\mathcal{A}$; triangles show the spectrum of $\mathcal{W}$, with blue for the cluster, purple for the steady state, and green for the topologically emerging state. Here, $\frac{\gamma_{\mathrm{in}}}{\gamma_{\mathrm{ex}}^{\prime}} = 8$ and $\frac{\gamma_{\mathrm{ex}}}{\gamma_{\mathrm{in}}^{\prime}} = 4$.
        (c) The topologically emerging state lies closer to the steady state in the stochastic spectrum within the topological regime, despite being the middle mode of the spectrum. As the non-reciprocity parameter $c$ increases, quantum clusters converge to one point, and stochastic clusters converge to two separate points, while the stochastic steady state remains fixed. Here $c$ is varied at fixed ratio $r = 0.8$, and $N = 6$.
        (d) The parameter $r$ is varied at fixed non-reciprocity $c = 0.8$. The gray shading indicates the topological phase, and the white shading indicates the trivial phase. The diagonal green line tracks the topologically emerging eigenmode, which approaches the steady state linearly as $r \to 1$. Here $N = 6$.}
	
	\label{fig:1dsshn1}
		\end{figure*}

Several key features are observed about the eigenvector structure as non-reciprocity increases. First, consistent with previous findings\cite{Nelson2024}, eigenvectors of different states in the cluster progressively overlap\label{fig:univec}, mirroring the collapse of the spectral cluster observed in the eigenvalue spectrum (Fig.~\ref{fig:lattice}(c)). 

Second, the analytical eigenvectors contain two distinct features that are visible both in the analytical expression and in Fig.~\ref{fig:lattice}(d): an exponential envelope and a sinusoidal modulation. The exponential envelope, proportional to $\left(\gamma_{\mathrm{in}}/\gamma_{\mathrm{in}}^{\prime}\right)^{(n-1)/2}$, governs the spatial localization of the modes typical of edge states, consistent with 
the non--hermitian skin effect 
\cite{yao2018edge,PhysRevLett.121.026808,PhysRevLett.123.066404}. At the same time, the eigenvectors retain a sinusoidal structure arising from the underlying lattice. These two features coexist (with the the exponential decay forming an envelope over the sinusoidal modulation).

Intriguingly, when system size $N$ is even, one of these states (highlighted in green in Fig.~\ref{fig:lattice}(d)) shows a step-like spatial profile due to the interference between the two sine terms (see details in Appendix.~\ref{sec:1dtes}). This state actually becomes increasingly long-lived in the topological phase, as we will see in the later sections that study topological models. Hence we dub this state the \emph{topologically emerging state} (TES), and will explore its properties more fully in Sec.~\ref{sec:pushed}. In this Hatano-Nelson chain, we note that at extreme $c$, the eigenvector localization of the stochastic spectral cluster completely overlaps with that of the TES, while the steady-state localization remains distinct (analytical derivation in Appendix.~\ref{sec:1dtes}). Together, these features demonstrate how increasing non-reciprocity reshapes both the spectral and spatial structure of stochastic eigenmodes.

\section{1D Non--Hermitian SSH Chain}
\label{sec:SSH}
The Su--Schrieffer--Heeger (SSH) model provides the simplest one-dimensional realization of nontrivial topology in quantum systems and therefore serves as a natural setting for exploring analogous topological phenomena in stochastic lattices \cite{jiang2020topological,lieu2018topological}. Physically, it describes a polyacetylene chain with alternating single and double bonds \cite{su1979solitons}. We extend the Hatano-Nelson chain considered above by relaxing the uniform-rate constraint and introducing alternating intracellular rates $\gamma_{\mathrm{in}}$ and extracellular rates $\gamma_{\mathrm{ex}}$, together with their backward counterparts $\gamma_{\mathrm{in}}^{\prime}$ and $\gamma_{\mathrm{ex}}^{\prime}$, as illustrated in Fig.~\ref{fig:1dsshn1}(a). Now, each unit cell $j$ consists of two sublattice sites $A$ and $B$. 

For the non--hermitian SSH chain, the quantum Hamiltonian takes the form \cite{kouachi2005}
\begin{equation}
\begin{aligned}
\mathcal{A}_{\mathrm{SSH}}
= \sum_{j=1}^{N}\,
&\gamma_{\mathrm{in}} \ket{j,B}\bra{j,A}
+ \gamma_{\mathrm{in}}' \ket{j,A}\bra{j,B}\\
&+\gamma_{\mathrm{ex}} \ket{j+1,A}\bra{j,B}
+\gamma_{\mathrm{ex}}' \ket{j,B}\bra{j+1,A}.
\end{aligned}
\label{eq:Hamiltonian}
\end{equation}

We study an open chain with $N$ sites characterized by alternating hopping rates \(\gamma_{\mathrm{in}},\gamma_{\mathrm{in}}'\) and \(\gamma_{\mathrm{ex}},\gamma_{\mathrm{ex}}'\). To simplify the large parameter space, we set 
\begin{equation}
\gamma_{\mathrm{in}}+\gamma_{\mathrm{ex}}'
=
\gamma_{\mathrm{ex}}+\gamma_{\mathrm{in}}',
\label{eq:ssh_diag_constraint}
\end{equation}
 which makes the non-corner diagonal elements for the stochastic operator uniform.

We then derive analytical results for both the quantum and the stochastic system (see details in Appendix \ref{sec:1Dsshanaspec}). For quantum system with odd size $N$, the eigenvalue problem reduces to a simple second-order recursion whose structure matches that of standard Chebyshev polynomials, so the spectrum can be obtained from their well-known zeros under open-boundary conditions. This leads to a set of symmetric bulk eigenvalues together with a single isolated boundary mode arising from the remaining block of the factorized determinant.

\begin{equation}
\begin{aligned}
\lambda_{A,k}^{\pm} &=
\pm \sqrt{
\gamma_{\mathrm{in}}\gamma_{\mathrm{in}}'
+\gamma_{\mathrm{ex}}\gamma_{\mathrm{ex}}'
+2\sqrt{
\gamma_{\mathrm{in}}\gamma_{\mathrm{in}}'
\gamma_{\mathrm{ex}}\gamma_{\mathrm{ex}}'
}\cos\!\left(\frac{2k\pi}{N+1}\right)
},\\
&\qquad k=1,\ldots,\frac{N-1}{2},\\[6pt]
\end{aligned}
\end{equation}

\begin{equation}
\lambda_A^{0}=0.
\end{equation}

For the stochastic system with even size $N$, a similarity transformation reduces the stochastic generator to a perturbed period-2-Toeplitz (a generalization of a Toeplitz matrix where the entries repeat with period 2) form whose characteristic polynomial factorizes into a Chebyshev-generated bulk sector and a quadratic boundary factor. The cluster eigenvalues follow from cosine quantization of the Chebyshev recurrence, while the boundary factor produces two isolated eigenvalues corresponding to the steady state and the topologically emerging state (TES).

We obtain the cluster spectrum as
\begin{equation}
\begin{aligned}
\lambda_{\mathcal{W}}^{\mathrm{upper}} &=
-(\gamma_{\mathrm{in}}+\gamma_{\mathrm{ex}}') \\
&+ 2\sqrt{
\gamma_{\mathrm{in}}\gamma_{\mathrm{in}}'
+ \gamma_{\mathrm{ex}}\gamma_{\mathrm{ex}}'
+ 2\sqrt{
\gamma_{\mathrm{in}}\gamma_{\mathrm{in}}'
\gamma_{\mathrm{ex}}\gamma_{\mathrm{ex}}'
}\cos\!\left(\dfrac{2k\pi}{N}\right)
},\\
& k=1,\dots,\frac{N}{2}-1.
\end{aligned}
\end{equation}

\begin{equation}
\begin{aligned}
&\lambda_{\mathcal{W}}^{\mathrm{lower}}=
-(\gamma_{\mathrm{in}}+\gamma_{\mathrm{ex}}')\\
&-2\sqrt{\gamma_{\mathrm{in}}\gamma_{\mathrm{in}}'
+\gamma_{\mathrm{ex}}\gamma_{\mathrm{ex}}'
+2\sqrt{\gamma_{\mathrm{in}}\gamma_{\mathrm{in}}'
\gamma_{\mathrm{ex}}\gamma_{\mathrm{ex}}'}
\cos\!\bigl(\tfrac{2(k-\frac{N}{2}+1)\pi}{N}\bigr)},\\
&k=\frac{N}{2},\dots,N-2.
\end{aligned}
\end{equation}

In addition, we verify the steady state as
\begin{equation}
\begin{split}
\lambda_{\mathrm{ss}}
&=
-(\gamma_{\mathrm{in}}-\gamma_{\mathrm{ex}}')\\[4pt]
&\quad
-\frac{
\gamma_{\mathrm{ex}}+\gamma_{\mathrm{ex}}'
-\sqrt{
(\gamma_{\mathrm{ex}}-\gamma_{\mathrm{ex}}')^2
+4\gamma_{\mathrm{in}}\gamma_{\mathrm{in}}'
}
}{2}\\
&=0.
\end{split}
\label{eq:lambdasteady_appendix}
\end{equation}

The topologically emerging state is
\begin{equation}
\begin{split}
\lambda_{\mathrm{TES}}
&=
-(\gamma_{\mathrm{in}}-\gamma_{\mathrm{ex}}')\\[4pt]
&\quad
-\frac{
\gamma_{\mathrm{ex}}+\gamma_{\mathrm{ex}}'
+\sqrt{
(\gamma_{\mathrm{ex}}-\gamma_{\mathrm{ex}}')^2
+4\gamma_{\mathrm{in}}\gamma_{\mathrm{in}}'
}
}{2}.
\end{split}
\label{eq:lambdates_appendix}
\end{equation}

Both systems share the same underlying three-term (Chebyshev-type) recurrence, but differ in boundary conditions: the odd quantum case matches standard Chebyshev initial conditions and admits closed-form roots, whereas the even stochastic case modifies the boundary structure, leading to factorization into bulk Chebyshev modes plus additional boundary-induced eigenvalues. 

By contrast, the even quantum case inherits the same recurrence but with shifted (non-Chebyshev) initial conditions that prevent reduction to a known polynomial family, so the spectrum is defined only implicitly and does not admit a simple closed-form solution.

\subsection{Neighboring size spectral correspondence requires even stochastic system size}

We ask whether the spectral width scales with system size differently than in the Hatano–Nelson chain studied previously. Solving for the spectrum analytically shows that the spectrum splits into two clusters (denoted above with $\pm$ superscripts for the quantum case and upper and lower superscripts for the stochastic case), whose widths approach asymptotic values as the system size increases. 

We next examine whether a system-size correspondence arises in this model. We find that the stochastic spectrum for a system of size \(N\) overlaps with that of the quantum system with \(N-1\) (Fig.~\ref{fig:1dsshn1}(b)). Unlike the Hatano-Nelson lattice, where this \(N \leftrightarrow N-1\) correspondence holds for all system sizes, in the SSH model it occurs only when \(N\) is even, reflecting the two-site unit-cell structure (Appendix~\ref{sec:1Dsshsize}).

\subsection{The topologically emerging state adds one more state near the steady state}
To accommodate the additional degrees of freedom in the 1D SSH lattice and to investigate the system’s behavior in the topological phase, we relax the constraint in Eq.~(\eqref{eq:ssh_diag_constraint}). Although the cluster eigenvalues can no longer be obtained analytically, numerical investigation reveals several robust features of the spectrum. In addition to the non-reciprocity ratio $c$ introduced earlier, we adopt a dimensionless parameter $r$ similar to previous papers \cite{tang2021topo,Nelson2024} that characterizes the relative strength of internal and external transition rates.
	\[
	\frac{\gamma_{\mathrm{ex}}}{\gamma_{\mathrm{in}}} = 
	\frac{\gamma_{\mathrm{ex}}^{\prime}}{\gamma_{\mathrm{in}}^{\prime}} = 
	\frac{r}{1 - r}, \qquad
	\frac{\gamma_{\mathrm{ex}}}{\gamma_{\mathrm{ex}}^{\prime}} = 
	\frac{\gamma_{\mathrm{in}}}{\gamma_{\mathrm{in}}^{\prime}} = 
	\frac{c}{1 - c}.
	\]
    
Here, $r$ controls the balance between internal and external couplings. For the Hermitian quantum SSH model, the spectrum undergoes a topological phase transition at $r=0.5$: when internal couplings dominate ($r<0.5$), the system is topologically trivial, while dominant external couplings ($r>0.5$) give rise to a topological phase characterized by zero-energy edge states \cite{su1979solitons,zhao2024interplay}.

We are interested in how non-reciprocity $c$ changes the quantum and stochastic spectrum. As \(c \to 1\) or \(c \to 0\), the quantum spectrum collapses from a finite-width cluster into a single point, as in the Hatano–Nelson chain (Fig.~\ref{fig:1dsshn1}(c)). In contrast, in the stochastic system each spectral cluster collapses to a single point, while the steady-state eigenvalue persists. Correspondingly, stochastic eigenstates whose eigenvalues overlap share the same localization profile (see Appendix~\ref{sec:sshtes}). Thus, non-reciprocity drives opposite spectral responses in the two systems: it concentrates quantum eigenvalues toward the zero-energy state, while pushing the stochastic clusters away from the steady state, thereby preserving a finite gap.
 
Overall, the stochastic SSH model retains several spectral features of the Hatano-Nelson lattice, while also exhibiting qualitatively new behavior absent in the quantum case. In particular, an additional long-lived mode, the topologically emerging state, appears and approaches the steady state, as shown in green in Fig.~\ref{fig:1dsshn1}(c), and is discussed in the next section.

	\subsection{The topologically emerging state becomes evident in the spectrum as the topological parameter is tuned}
    \label{sec:pushed}

We identify a special eigenmode of the stochastic generator, which we term the \emph{topologically emerging state} (TES). In the topological phase, as $r \to 1$, this mode separates from the rest of the spectrum and approaches the steady-state eigenvalue, signaling the emergence of a long-lived state that is absent in the trivial phase. The TES eigenvalue is independent of the non-reciprocity parameter $c$. It moves linearly toward the steady-state eigenvalue as $r$ increases, i.e., 
\begin{equation}
\lambda_{\mathrm{TES}} = 2r - 2.
\end{equation}

This expression is true for the 1D SSH chain but also in the 1D non--hermitian Hatano-Nelson chain when $r=0.5$. We will see that this TES persists across a broad class of non--hermitian stochastic lattices, including the 1D Hatano-Nelson chain, the 1D SSH chain, and a 2D SSH model (Sec.~\ref{sec:2D}). 

\begin{figure}
	\centering
	\includegraphics{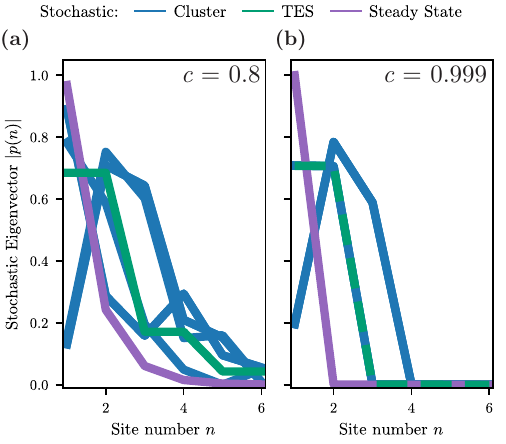}
	\caption{\textbf{The step-like spatial profile of TES persisted in 1D SSH model.}
		Left: Eigenvector profiles at $c = 0.8$, $r = 0.8$ for $N = 6$. 
        Right: Eigenvector profiles at $c = 0.999$, $r = 0.8$ for $N = 6$. Same color scheme as Fig.~\ref{fig:lattice}(d). 
		}
	\label{fig:sshvec}
	
\end{figure}

Notably, while the quantum spectrum remains symmetric in both the trivial and topological regimes, the stochastic spectrum develops an increasing number of long-lived modes because of the contribution of TES and other states in the strong topological limit (Fig.~\ref{fig:1dsshn1}(d)). This behavior highlights a fundamental distinction between quantum and stochastic topological systems: in stochastic systems, entering the topological phase with increasing non-reciprocity  generically promotes spectral crowding near the steady state rather than protecting isolated edge modes.
\begin{figure}
	\centering
	\includegraphics{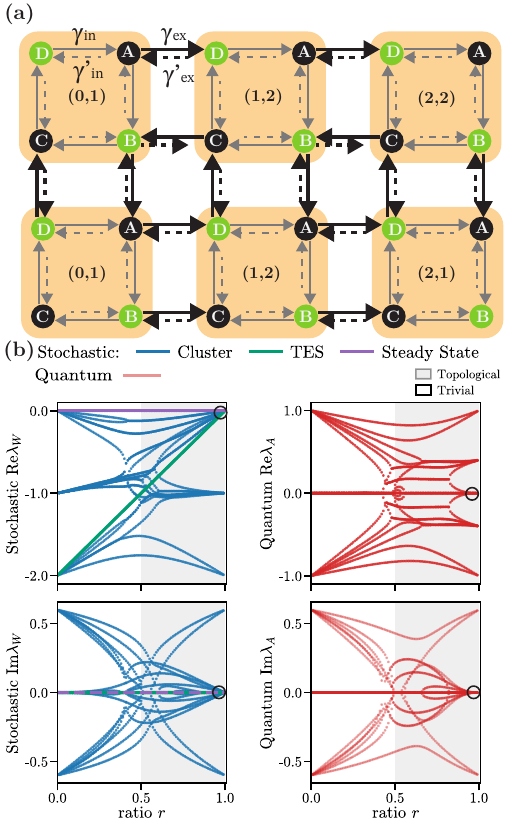}[ht!]
	\caption{\textbf{Properties of 2D SSH non--hermitian quantum and stochastic lattices.}
        (a) A 2D lattice with non-reciprocal transition rates, parameterized by alternating forward rates $\gamma_{\mathrm{in}}$, $\gamma_{\mathrm{ex}}$ and backward rates $\gamma_{\mathrm{in}}^{\prime}$, $\gamma_{\mathrm{ex}}^{\prime}$. The strength of non-reciprocity is characterized by $c$, and the relative strength of internal and external rates is characterized by $r$. The yellow shading indicates one unit cell. A pair of numbers $(i,j)$ indicates the position of the cell. 
        (b) We show the real and imaginary spectra of the quantum and stochastic systems, where gray shading indicates the topological phase and the white shading indicates the trivial phase. In the strongly topological regime, the topologically emerging state (green line) and other states cause the stochastic spectrum (blue lines) to contain a significantly larger number of states near the steady state (purple line), as compared to the the quantum spectrum (red lines) near zero energy. These points of comparison are indicated with circles in both the stochastic and quantum systems, in the strong topological limit. Here, $c = 0.8$, $N_i = 2$, and $N_j = 3$.}
	\label{fig:2dsshn1}
\end{figure}

\subsection{Analytical eigenvector of the topologically emerging state}

In the 1D SSH chain, with lattice sites labeled by \( n = 1,\ldots,N \), the localization of the topologically emerging state (Fig.~\ref{fig:sshvec}) and the steady state is independent of $r$ . The corresponding eigenvector of TES is
\begin{equation}
p^{(\mathrm{TES})}_n
=
(-1)^{n}
\left(\frac{c}{c-1}\right)^{\frac{N}{2}-1-\left\lfloor \frac{n-1}{2} \right\rfloor},
\qquad
n = 1,\ldots,N.
\label{eq:TES_eigenvector}
\end{equation}

The eigenvector of the stochastic steady state is 
\begin{equation}
p^{(\mathrm{steady})}_n
=
(-1)^{n-N-1}
\left(\frac{c}{c-1}\right)^{\,N-n+1},
\qquad
n = 1,\ldots,N.
\end{equation}

The eigenvectors of TES and stochastic steady state are mathematically similar in form because both eigenvectors are built from the same underlying geometric decay set by the non-reciprocity ratio 
$\frac{\gamma_{\mathrm{in}}}{\gamma_{\mathrm{in}}^{\prime}}$, or $\frac{c}{c-1}$, combined with an alternating sign pattern that reflects the two-sublattice structure of the chain.

The steady-state vector applies this geometric factor uniformly across all sites, giving a smooth exponential profile along the lattice, whereas the TES groups sites in pairs through the floor function in the exponential of Eq.~\eqref{eq:TES_eigenvector}. The floor function will take $\frac{n-1}{2}$ and round it down to the nearest integer, producing a step-like version of the same envelope. In other words, the TES can be viewed as a coarse-grained or sublattice-modulated version of the steady state, with the same localization length but a different internal structure.

Despite residing in the cluster of the spectrum, the topologically emerging state exhibits a distinctive step-like spatial profile, qualitatively different from conventional edge or cluster modes for both Hatano-Nelson chain (Fig.~\ref{fig:lattice}(c)) and the SSH chain (Fig.~\ref{fig:sshvec}). We find that the characteristic wavelength is two sites in both chains. The analytical derivations are in Appendix.~\ref{sec:1dtes} and Appendix.~\ref{sec:sshtes}.

Overall, the localization profile of the stochastic SSH chain sharply contrasts with the quantum SSH chain, where in previous literature, it was shown all zero-energy states are edge localized\cite{Nelson2024}. Here, in contrast, stochastic systems exhibit zero-eigenvalue states that are not necessarily edge-localized at $c=0.5$, as exemplified by the topologically emerging state (TES). How the localization properties of the topologically emerging state depend jointly on $r$ and $c$ in higher-dimensional non-Hermitian SSH lattices is examined in the following section.

	\section{2D Non--Hermitian SSH Lattice}
    \label{sec:2D}

To probe whether the topologically emerging mode identified in 1D is tied to the specific geometry of the chain or reflects a more general feature of the stochastic systems, we extend the SSH construction to two dimensions, where steady states can support circulating edge currents as in the previously proposed model \cite{tang2021topo,Nelson2024}.

Using the same parameterization with non-reciprocity $c$ and the ratio $r$, we construct a non--hermitian SSH lattice on an $N_i \times N_j$ grid with four sites ($A,B, C, D$) per unit cell. Each unit cell at position $\mathbf{r}=(i,j)$ contains intracell couplings $\gamma_{\mathrm{in}}$ and $\gamma_{\mathrm{in}}'$, together with intercell couplings $\gamma_{\mathrm{ex}}$ and $\gamma_{\mathrm{ex}}'$ connecting neighboring unit cells along the $\hat{i}$ and $\hat{j}$ directions.

The corresponding quantum operator is given in Appendix \ref{sec:2DSSHana}. Non-reciprocity is introduced through asymmetric forward and backward hopping amplitudes, allowing probability or amplitude flow to be directionally biased in both spatial directions (though taken to be identical along the $\hat{i}$ and $\hat{j}$ directions).

\subsection{Topology drives opposite spectral accumulation patterns in quantum and stochastic systems}

Increasing topology concentrates the stochastic spectrum near the steady state. In the 2D SSH lattice, as the ratio $r$ increases, the stochastic spectrum exhibits pronounced eigenvalue splitting, signaling the separation of long-lived modes from the cluster of rapidly decaying states (Fig.~\ref{fig:2dsshn1}(b)). This splitting reflects an increasing spectral organization: as the system is driven deeper into the topological regime, slow modes become progressively isolated from the fast-relaxing cluster. In the large-$r$ limit, the eigenvalues further collapse into isolated points in the complex plane, with an increasing number of modes clustering near the steady state, leading to a highly concentrated spectrum of slow dynamics. Together, these features indicate that increasing topology promotes the accumulation of long-lived modes near the steady state. 

We compare this behavior with the quantum system in the same topological limit. Far fewer states in the quantum system accumulate at the zero-energy state (Fig.~\ref{fig:2dsshn1}(b)), as the clusters are instead pushed away and the gap around the zero-energy mode increases. This contrast demonstrates that topology drives opposite spectral responses in the two systems. Stochastic spectra become increasingly concentrated near the steady state, whereas quantum spectra become increasingly separated from the zero-energy state.

\subsection{Topologically emerging state persisting in 2D}

In the strongly topological regime, the stochastic spectrum acquires an additional eigenstate near the steady state, corresponding to the topologically emerging state. We examine the spectral properties of the topologically emerging state in two dimensions and their dependence on system parameters. As in the 1D SSH lattice, its eigenvalue scales linearly as $2r - 2$ and remains independent of the $c$, indicating that its existence and robustness are governed primarily by the overall ratio $r$ rather than non-reciprocity (Fig.~\ref{fig:2dsshn1}(b)). Using the same parameterization, we find that the key spectral features of the TES identified in one dimension persist in the presence of circulating currents.

\section{Conclusion}
In this work, we compare the spectra of quantum and stochastic systems defined on identical lattice networks. By deriving analytical spectra for both quantum Hamiltonians and stochastic generators, we identify control parameters governing edge localization, relaxation timescales, and spectral gap changes under non-reciprocity. Starting from a one-dimensional non-reciprocal chain, we show that increasing non-reciprocity drives quantum eigenvalues toward the zero-energy state, whereas stochastic eigenvalues are pushed away from the steady state. 
Thus, non-reciprocity drives opposite spectral responses in the two systems, concentrating quantum eigenvalues while repelling stochastic modes from the steady state. This contrast reflects probability conservation, which fixes the steady-state eigenvalue and reshapes the surrounding spectrum. 

While the quantum system supports topologically protected edge modes, the stochastic system develops an enhanced density of eigenvalues near the steady state and an increasing number of long-lived modes in the strong topological limit. The quantum spectra move away from the zero-energy state, and the stochastic spectra become increasingly concentrated near the steady state. In particular, a topologically emerging state (TES) appears as an additional eigenmode moving linearly toward the steady state in the stochastic system, providing a distinct mechanism for generating long-lived modes. Thus, in contrast to quantum topology, which protects edge states through spectral gaps, stochastic topology manifests through a redistribution of relaxation timescales and the emergence of slower modes.

Although introduced for analytical tractability, the framework may also inform stochastic systems in biology and ecology, where persistence and timescales are central concerns \cite{chave2004neutral,rosindell2012case}. 
Taken together, these results show that stochastic systems can exhibit topology reminiscent of quantum systems while obeying distinct spectral principles. Non-reciprocity and topology therefore act as complementary control parameters that generate long-lived modes and reshape spectral gaps in fundamentally different ways in quantum and stochastic systems. By revealing how these parameters control spectral structure, this work provides analytical tools for designing and controlling non-equilibrium stochastic transport, biochemical networks, and active systems. 

In this context, the spectral differences we identify may offer insight into how control strategies differ between quantum-inspired and stochastic network dynamics. 
More broadly, our results may also inform network control frameworks, where the distinction between adjacency-like and Laplacian operators \cite{glasser1985control,yuksel2013stochastic} plays a central role in determining controllability and dynamical response.

\section{Acknowledgments}
We thank Tolya Kolomeisky for the idea to start comparing the spectra of these systems, and Chongbin Zheng for discussions and help. This work was supported by a Dessler Scholarship from the Department of Physics and Astronomy at Rice University, as well as by the NSF Center for Theoretical Biological Physics (PHY-2019745), the NSF CAREER Award (DMR-2238667), and the Chan Zuckerberg Initiative.

\appendix
\section{Analytical expression for the 1D Hatano-Nelson chain spectrum}
\label{sec:1Danaspec}

The quantum Hamiltonian $\mathcal{A}$ takes the form of a tridiagonal Toeplitz matrix, while the corresponding stochastic generator $\mathcal{W}$ is a corner-perturbed tridiagonal Toeplitz matrix\cite{noschese2013tridiagonal}. 

In the bulk, these matrices have constant off-diagonal elements and diagonal terms, so their eigenvalue problem reduces to a second-order linear difference equation\cite{alvarez2012spectral}.

Following Ref.~\cite{alvarez2012spectral}, we derive the result. Let $v_j$ denote the $j$th component of an eigenvector. For a uniform tridiagonal matrix with off-diagonal entries $\gamma_{\mathrm{in}}$ and $\gamma_{\mathrm{in}}^{\prime}$, the eigenvalue equation in the bulk reads
\begin{equation}
\gamma_{\mathrm{in}}^{\prime} v_{j-1}
+ \gamma_{\mathrm{in}} v_{j+1}
= \lambda v_j .
\end{equation}

This is a homogeneous second-order recurrence relation. 
Seeking solutions of the form $v_j = y^j$ yields the characteristic equation
\begin{equation}
\gamma_{\mathrm{in}} y^2 - \lambda y + \gamma_{\mathrm{in}}^{\prime} = 0.
\end{equation}

Introducing the rescaled eigenvalue
\begin{equation}
x = \frac{\lambda}{2\sqrt{\gamma_{\mathrm{in}}\gamma_{\mathrm{in}}^{\prime}}},
\end{equation}
the characteristic equation becomes
\begin{equation}
y + y^{-1} = 2x.
\end{equation}
Thus, the general solution in the bulk can be written as
\begin{equation}
v_j = A y_+^{\,j} + B y_-^{\,j},
\end{equation}
where $y_\pm = x \pm \sqrt{x^2-1}$.

Defining $x=\cos\theta$, we obtain
\begin{equation}
y_\pm = e^{\pm i\theta}.
\end{equation}
The bulk solution therefore reduces to
\begin{equation}
v_j = C \sin(j\theta),
\end{equation}
after imposing appropriate boundary conditions. 
The resulting recurrence relation for the characteristic polynomial coincides with that of the Chebyshev polynomials of the second kind,
\begin{equation}
U_N(x) = \frac{\sin[(N+1)\arccos x]}{\sin(\arccos x)},
\end{equation}
which satisfy
\begin{equation}
U_{n+1}(x) = 2x U_n(x) - U_{n-1}(x).
\end{equation}

Hence the eigenvalues are determined by the zeros of $U_N(x)$.

For the quantum Hamiltonian $\mathcal{A}$, open boundary conditions impose
\begin{equation}
\theta_k = \frac{k\pi}{N+1}, 
\qquad k=1,2,\dots,N,
\end{equation}
leading to eigenvalues
\begin{equation}
\lambda_\mathcal{A}
= -2\sqrt{\gamma_{\mathrm{in}}\gamma_{\mathrm{in}}^{\prime}}
\cos\!\left(\frac{k\pi}{N+1}\right).
\end{equation}

For the stochastic generator $\mathcal{W}$, probability conservation imposes a zero-sum constraint on each column, which shifts the diagonal by 
$-(\gamma_{\mathrm{in}}+\gamma_{\mathrm{in}}^{\prime})$. 
The bulk recurrence remains the same, but the boundary condition changes, yielding
\begin{equation}
\theta_k = \frac{k\pi}{N}, 
\qquad k=1,2,\dots,N-1.
\end{equation}
Therefore the non-steady eigenvalues are
\begin{equation}
\lambda_\mathcal{W}
= -(\gamma_{\mathrm{in}}+\gamma_{\mathrm{in}}^{\prime})
+ 2\sqrt{\gamma_{\mathrm{in}}\gamma_{\mathrm{in}}^{\prime}}
\cos\!\left(\frac{k\pi}{N}\right).
\end{equation}

In addition, probability conservation guarantees a steady-state eigenvalue
\begin{equation}
\lambda_{\mathrm{ss}} = 0.
\end{equation}

\section{Characteristic wavelength of the TES eigenvector in the 1D Hatano-Nelson chain}
\label{sec:1dtes}

\begin{figure}
	\centering
	\includegraphics{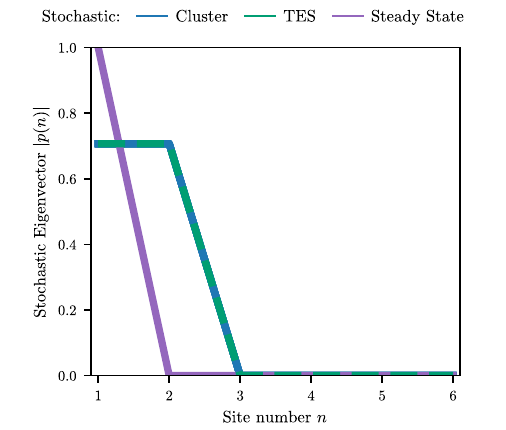}
	\caption{\textbf{At extreme non-reciprocity, the eigenvector of the cluster completely overlaps with that of the TES.}
        Eigenvector profiles at $c = 0.99$, $r = 0.5$ for $N = 6$. Same color scheme as Fig.~\ref{fig:lattice}(d).  
		} 
	\label{fig:univec}
	
\end{figure}

TES eigenvector corresponds to the eigenvalue in the middle of the spectrum. Setting $k = N/2$ in the cluster eigenvector expression [Eq.~\eqref{eq:cluster_eigenvector}] gives
\begin{equation}
p^{(N/2)}_{\mathrm{cluster},n}
=
\left(\frac{\gamma_{\mathrm{in}}}{\gamma'_{\mathrm{in}}}\right)^{\frac{n-1}{2}}
\left[
\sin\!\left(\frac{n\pi}{2}\right)
-
\sqrt{\frac{\gamma'_{\mathrm{in}}}{\gamma_{\mathrm{in}}}}
\sin\!\left(\frac{(n-1)\pi}{2}\right)
\right].
\end{equation}

The corresponding effective wave number is
\begin{equation}
q = \frac{k\pi}{N} = \frac{\pi}{2},
\end{equation}
which implies a characteristic wavelength
\begin{equation}
\lambda = \frac{2\pi}{q} = 4.
\end{equation}

Equivalently,
\begin{equation}
\sin\!\left(\frac{nk\pi}{N}\right)
=
\sin\!\left(\frac{n\pi}{2}\right),
\end{equation}
which repeats every four lattice sites. The $k=N/2$ cluster eigenvector therefore exhibits a four-site periodic structure modulated by the envelope
\(
(\gamma_{\mathrm{in}}/\gamma'_{\mathrm{in}})^{(n-1)/2}.
\)

At $q=\pi/2$, the sine functions take only the discrete values
\begin{equation}
\sin\!\left(\frac{n\pi}{2}\right)
=
\{0,1,0,-1,0,1,\ldots\},
\end{equation}
\begin{equation}
\sin\!\left(\frac{(n-1)\pi}{2}\right)
=
\{-1,0,1,0,-1,0,\ldots\}
\end{equation}
Thus at each lattice site only one of the two sine terms is nonzero.

Let $n=2m$ (even), then
\begin{equation}
\sin\!\left(\frac{n\pi}{2}\right)=0,
\qquad
\sin\!\left(\frac{(n-1)\pi}{2}\right)=(-1)^{m+1},
\end{equation}
which yields
\begin{equation}
p_{even}
=
(-1)^m
\sqrt{\frac{\gamma'_{\mathrm{in}}}{\gamma_{\mathrm{in}}}}
\left(\frac{\gamma_{\mathrm{in}}}{\gamma'_{\mathrm{in}}}\right)^{m-\frac12}.
\end{equation}

For $n=2m+1$ (odd),
\begin{equation}
\sin\!\left(\frac{n\pi}{2}\right)=(-1)^m,
\qquad
\sin\!\left(\frac{(n-1)\pi}{2}\right)=0,
\end{equation}
so that
\begin{equation}
p_{2m+1}
=
(-1)^m
\left(\frac{\gamma_{\mathrm{in}}}{\gamma'_{\mathrm{in}}}\right)^m.
\end{equation}

Using
\begin{equation}
\sqrt{\frac{\gamma'_{\mathrm{in}}}{\gamma_{\mathrm{in}}}}
\left(\frac{\gamma_{\mathrm{in}}}{\gamma'_{\mathrm{in}}}\right)^{m-\frac12}
=
\left(\frac{\gamma_{\mathrm{in}}}{\gamma'_{\mathrm{in}}}\right)^m,
\end{equation}
we obtain
\begin{equation}
p_{2m} = p_{2m+1}
=
(-1)^m
\left(\frac{\gamma_{\mathrm{in}}}{\gamma'_{\mathrm{in}}}\right)^m. 
\end{equation}

Thus the amplitudes are identical on neighboring pairs of sites,
\begin{equation}
(p_{0},p_{1}),\ (p_{2},p_{3}),\ (p_{4},p_{5}),\ \ldots ,
\end{equation}
while changing only when $m$ increases. The eigenvector therefore forms two-site plateaus whose magnitude varies geometrically with $m$. Moreover, at extreme non-reciprocity, the eigenvectors of the cluster completely overlap with the eigenvector of the TES (Fig.~\ref{fig:univec}).

\section{Analytical expression for the 1D SSH chain spectrum}
\label{sec:1Dsshanaspec}
\subsection{Analytical expression for odd-sized quantum system}
Adapting the mathematic approach of Ref.\cite{kouachi2008eigenvalues}, we consider an open chain with an odd number of sites, \(N\), and alternating hopping rates
\(\gamma_{\mathrm{in}},\gamma_{\mathrm{in}}'\) and \(\gamma_{\mathrm{ex}},\gamma_{\mathrm{ex}}'\).
To isolate the effect of alternating couplings, we impose Eq.\eqref{eq:ssh_diag_constraint}
so that all diagonal entries except for the two corners of the stochastic generator \(\mathcal W\) are equal.
This removes any site-dependent on-site shift and reduces the problem to a tridiagonal
\(2\)-Toeplitz form with constant bulk diagonal and alternating off-diagonal couplings. For the quantum system, we consider the characteristic equation
\begin{equation}
\det(\mathcal{A}-\lambda I)=0,
\label{eq:char_eq_appendix}
\end{equation}
where \(\mathcal{A}\) is the tridiagonal matrix associated with the open chain. 

Writing $\mathcal{A} - \lambda I \equiv S$ in block form and applying the recursive identity, where $\alpha^{(k)}$ denotes the matrix obtained after $k$ steps of block Gaussian elimination, with
\begin{equation}
\alpha_{ij}^{(0)} = S_{ij},
\end{equation}

\begin{equation}
\alpha_{ij}^{(k+1)}
=
\alpha_{ij}^{(k)}
-
\alpha_{i,N-k}^{(k)}
\bigl(\alpha_{N-k,N-k}^{(k)}\bigr)^{-1}
\alpha_{N-k,j}^{(k)},
\label{eq:alpha_recursion_appendix}
\end{equation}
the determinant factorizes into the product of the diagonal Schur-complement blocks \cite{powell2011calculating}.

For the alternating chain, the recursion closes into a second-order difference equation. After imposing the diagonal constraint Eq.\ref{eq:ssh_diag_constraint}, the bulk recursion takes the form
\begin{equation}
x_{j+1}
=
\frac{\lambda^2-\gamma_{\mathrm{in}}\gamma_{\mathrm{in}}'
-\gamma_{\mathrm{ex}}\gamma_{\mathrm{ex}}'}
{\sqrt{\gamma_{\mathrm{in}}\gamma_{\mathrm{in}}'
\gamma_{\mathrm{ex}}\gamma_{\mathrm{ex}}'}}
\,x_j
-
x_{j-1}.
\label{eq:bulk_recursion_appendix}
\end{equation}
Introducing
\begin{equation}
\alpha
=
\frac{\lambda^2-\gamma_{\mathrm{in}}\gamma_{\mathrm{in}}'
-\gamma_{\mathrm{ex}}\gamma_{\mathrm{ex}}'}
{\sqrt{\gamma_{\mathrm{in}}\gamma_{\mathrm{in}}'
\gamma_{\mathrm{ex}}\gamma_{\mathrm{ex}}'}},
\label{eq:alpha_def_appendix}
\end{equation}
Eq.~\eqref{eq:bulk_recursion_appendix} becomes 
\begin{equation}
x_{j+1}=\alpha x_j-x_{j-1}.
\label{eq:chebyshev_recursion_appendix}
\end{equation}
This is the standard Chebyshev-type recursion, whose oscillatory solutions are obtained by setting
\begin{equation}
\alpha_k = 2\cos q_k.
\label{eq:alpha_q_appendix}
\end{equation}
For an open chain, the boundary conditions quantize the allowed momenta as
\begin{equation}
q_k=\frac{2k\pi}{N+1},
\qquad
k=1,\dots,\frac{N-1}{2},
\label{eq:qk_appendix}
\end{equation}
so that
\begin{equation}
\alpha_k=2\cos\!\left(\frac{2k\pi}{N+1}\right).
\label{eq:alphak_appendix}
\end{equation}
Substituting Eq.~\eqref{eq:alphak_appendix} into Eq.~\eqref{eq:alpha_def_appendix} gives
\begin{equation}
\lambda^2
=
\gamma_{\mathrm{in}}\gamma_{\mathrm{in}}'
+\gamma_{\mathrm{ex}}\gamma_{\mathrm{ex}}'
+\sqrt{\gamma_{\mathrm{in}}\gamma_{\mathrm{in}}'
\gamma_{\mathrm{ex}}\gamma_{\mathrm{ex}}'}\,\alpha_k .
\label{eq:lambda2_appendix}
\end{equation}
Hence the bulk branches are
\begin{equation}
\lambda_k^{(\pm)}
=
\pm
\sqrt{
\gamma_{\mathrm{in}}\gamma_{\mathrm{in}}'
+\gamma_{\mathrm{ex}}\gamma_{\mathrm{ex}}'
+\sqrt{
\gamma_{\mathrm{in}}\gamma_{\mathrm{in}}'
\gamma_{\mathrm{ex}}\gamma_{\mathrm{ex}}'}\,\alpha_k
},
\label{eq:bulk_branches_appendix}
\end{equation}

\begin{equation}
k=1,\dots,\frac{N-1}{2}.
\end{equation}

In addition to these bulk solutions, the recursion leaves one isolated root,
\begin{equation}
\lambda_{\mathcal{A}}^{0}=0.
\label{eq:isolated_root_appendix}
\end{equation}
which follows from the remaining one-dimensional Schur-complement block after the bulk sector has been factorized.

\subsection{Analytical expression for even-sized stochastic system}
Here for the stochastic matrix, we consider an open chain with an even number of sites, \(N\), and alternating hopping rates
\(\gamma_{\mathrm{in}},\gamma_{\mathrm{in}}'\) and \(\gamma_{\mathrm{ex}},\gamma_{\mathrm{ex}}'\).
We impose Eq.\eqref{eq:ssh_diag_constraint}
so that all diagonal entries except for the two corners of the stochastic generator \(\mathcal W\) are equal.

A standard similarity transformation can be used to absorb the non-reciprocal amplitudes
into the basis vectors, so that the spectrum depends only on the products
\begin{equation}
d_1^2=\gamma_{\mathrm{in}}\gamma_{\mathrm{in}}',
\qquad
d_2^2=\gamma_{\mathrm{ex}}\gamma_{\mathrm{ex}}'.
\label{eq:d1d2defs}
\end{equation}

After this transformation, the eigenvalue problem reduces to the characteristic equation
of a perturbed tridiagonal \(2\)-Toeplitz matrix. For even \(N\), the characteristic
polynomial factorizes into a bulk part and a boundary part,
\begin{equation}
\chi_N(\lambda)=Q(\lambda)\,P_{m-1}(\lambda),
\label{eq:char_factorization}
\end{equation}
where \(P_{m-1}\) generates the bulk spectral branches and \(Q(\lambda)\) gives two
additional isolated eigenvalues.

Eliminating one sublattice gives a second-order recurrence relation for amplitudes on the
remaining sublattice,
\begin{equation}
\psi_{n+1}+\psi_{n-1}
=
2\cos\theta\,\psi_n,
\label{eq:bulk_recurrence}
\end{equation}
with
\begin{equation}
\cos\theta
=
\frac{S-(d_1^2+d_2^2)}{2d_1d_2},
\qquad
S \equiv \frac{1}{4}\bigl(\lambda+\gamma_{\mathrm{in}}+\gamma_{\mathrm{ex}}'\bigr)^2
\label{eq:S_theta_relation}.
\end{equation}

Equation~\eqref{eq:bulk_recurrence} is the Chebyshev recurrence, so the bulk solutions are
generated by Chebyshev polynomials of the second kind. Imposing open-boundary conditions
quantizes the phase as
\begin{equation}
\theta_k=
\begin{cases}
\dfrac{2k\pi}{N}, & k=1,\dots,\frac{N-2}{2},\\[6pt]
\dfrac{2(k-\frac{N}{2}+1)\pi}{N}, & k=\frac{N}{2},\dots,N-2.
\end{cases}
\label{eq:thetak_appendix}
\end{equation}

Substituting \eqref{eq:thetak_appendix} into \eqref{eq:S_theta_relation} yields
\begin{equation}
\begin{split}
S_k
&= d_1^2 + d_2^2 + 2 d_1 d_2 \cos \theta_k \\
&= \gamma_{\mathrm{in}}\gamma_{\mathrm{in}}'
   + \gamma_{\mathrm{ex}}\gamma_{\mathrm{ex}}'
   + 2\sqrt{
     \gamma_{\mathrm{in}}\gamma_{\mathrm{in}}'
     \gamma_{\mathrm{ex}}\gamma_{\mathrm{ex}}'
   }\cos\theta_k.
\end{split}
\label{eq:Sk_appendix}
\end{equation}

Therefore the \(N-2\) bulk eigenvalues are
\begin{equation}
\lambda_k=
\begin{cases}
-(\gamma_{\mathrm{in}}+\gamma_{\mathrm{ex}}')+2\sqrt{S_k},
& k=1,\dots,\frac{N-2}{2},\\[6pt]
-(\gamma_{\mathrm{in}}+\gamma_{\mathrm{ex}}')-2\sqrt{S_k},
& k=\frac{N}{2},\dots,N-2.
\end{cases}
\label{eq:bulk_eigs_appendix}
\end{equation}

Thus the spectrum separates into two branches, corresponding to the two signs in
Eq.~\eqref{eq:bulk_eigs_appendix}, exactly as in the usual SSH problem.

The remaining two eigenvalues do not come from the Chebyshev sector. Instead, they are the
roots of the quadratic boundary factor \(Q(\lambda)\) in Eq.~\eqref{eq:char_factorization},
\begin{equation}
\begin{split}
Q(\lambda)
&=
\bigl[\lambda+(\gamma_{\mathrm{in}}-\gamma_{\mathrm{ex}}')\bigr]^2 \\
&\quad
+(\gamma_{\mathrm{ex}}+\gamma_{\mathrm{ex}}')
\bigl[\lambda+(\gamma_{\mathrm{in}}-\gamma_{\mathrm{ex}}')\bigr] \\
&\quad
+\gamma_{\mathrm{ex}}\gamma_{\mathrm{ex}}'
-\gamma_{\mathrm{in}}\gamma_{\mathrm{in}}'.
\end{split}
\label{eq:Qlambda_appendix}
\end{equation}

Solving \(Q(\lambda)=0\) gives
\begin{equation}
\lambda_{\pm}
=
-(\gamma_{\mathrm{in}}-\gamma_{\mathrm{ex}}')
-\frac{
\gamma_{\mathrm{ex}}+\gamma_{\mathrm{ex}}'
\pm
\sqrt{
(\gamma_{\mathrm{ex}}-\gamma_{\mathrm{ex}}')^2
+4\gamma_{\mathrm{in}}\gamma_{\mathrm{in}}'
}
}{2}.
\label{eq:lambdapm_appendix}
\end{equation}

These are precisely the two isolated eigenvalues outside the bulk branches.

For a stochastic generator, one of these roots must be the steady state, so probability
conservation fixes
\begin{equation}
\lambda_{\mathrm{ss}}=0.
\label{eq:ss_zero_appendix}
\end{equation}
Accordingly, we verified 
\begin{equation}
\begin{split}
\lambda_{\mathrm{ss}}
&=
-(\gamma_{\mathrm{in}}-\gamma_{\mathrm{ex}}')\\[4pt]
&\quad
-\frac{
\gamma_{\mathrm{ex}}+\gamma_{\mathrm{ex}}'
-\sqrt{
(\gamma_{\mathrm{ex}}-\gamma_{\mathrm{ex}}')^2
+4\gamma_{\mathrm{in}}\gamma_{\mathrm{in}}'
}
}{2}=0,
\end{split}
\label{eq:lambdasteady_appendix}
\end{equation}
and the other root as the topologically emerging state,
\begin{equation}
\begin{split}
\lambda_{\mathrm{TES}}
&=
-(\gamma_{\mathrm{in}}-\gamma_{\mathrm{ex}}')\\[4pt]
&\quad
-\frac{
\gamma_{\mathrm{ex}}+\gamma_{\mathrm{ex}}'
+\sqrt{
(\gamma_{\mathrm{ex}}-\gamma_{\mathrm{ex}}')^2
+4\gamma_{\mathrm{in}}\gamma_{\mathrm{in}}'
}
}{2}.
\end{split}
\label{eq:lambdates_appendix}
\end{equation}

In summary, the open-chain \(1\)D SSH generator with uniform diagonal
constraint Eq.~\eqref{eq:ssh_diag_constraint} has \(N-2\) bulk eigenvalues given by
Eq.~\eqref{eq:bulk_eigs_appendix} and two boundary-induced eigenvalues given by
Eq.~\eqref{eq:lambdasteady_appendix} and Eq.~\eqref{eq:lambdates_appendix}. The former are
set by the Chebyshev quantization of the \(2\)-Toeplitz bulk recurrence, while the latter
arise from the quadratic boundary factor in the full characteristic polynomial.

\begin{figure}
	\centering
	\includegraphics{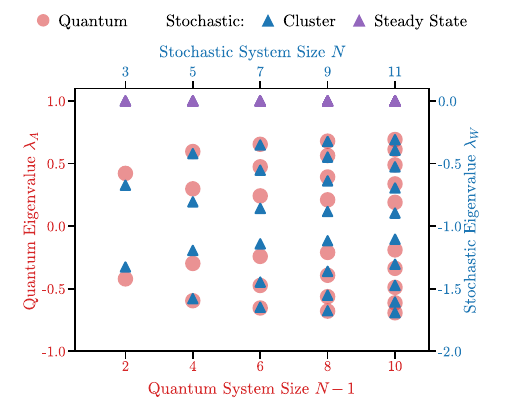}
	\caption{\textbf{Spectral correspondence between odd-sized stochastic systems and even-sized quantum systems.}
        The half-integer–cell stochastic cluster of system size $N$ does not completely overlap with the quantum cluster of size $N{-}1$. Blue triangles denote the stochastic system (size $N$, odd), purple triangles denote the stochastic steady state, and red circles denote the quantum system. Here $\frac{\gamma_{\mathrm{in}}}{\gamma_{\mathrm{ex}}^{\prime}} = 8$ and $\frac{\gamma_{\mathrm{ex}}}{\gamma_{\mathrm{in}}^{\prime}} = 4$.}
	\label{fig:sshodd}
	
\end{figure}

\subsection{Relationship between even-sized and odd-sized quantum system}
We did not derive the analytical solution for quantum system with even $N$ size because obtaining closed-form eigenvalues is significantly more difficult for even system size. The origin of this difficulty can be understood from the structure of the characteristic polynomial\cite{gover1994eigenproblem}. In both even and odd $N$ cases, the determinant reduces to a three-term recursion of the form
\begin{equation}
P_{k+1}(x) = x\,P_k(x) - P_{k-1}(x).
\label{eq:cheb_rec}
\end{equation}

For odd system size \(N\), the initial conditions
\begin{equation}
P_0(x)=1,\qquad P_1(x)=x,
\label{eq:cheb_init}
\end{equation}
match those of the Chebyshev polynomials of the first kind, \(T_k(x)\), defined by
\begin{equation}
T_k(\cos \theta) = \cos(k\theta).
\end{equation}
Therefore,
\begin{equation}
P_k(x)=T_k(x).
\end{equation}

The eigenvalues follow from the well-known zeros
\begin{equation}
x_k = 2\cos\!\left(\frac{2k\pi}{N+1}\right), \qquad k=1,\dots,\frac{N-1}{2}.
\label{eq:cheb_roots}
\end{equation}

In contrast, for even system size \(N=2m\), although the same recursion Eq.\eqref{eq:cheb_rec} still holds, the initial conditions are modified,
\begin{equation}
P_0(x)=1,\qquad P_1(x)=x-c,
\label{eq:shifted_init}
\end{equation}
where \(c\) depends on the off-diagonal parameters. As a result, the resulting polynomials are no longer Chebyshev polynomials, and their zeros are not given by a simple cosine form. The eigenvalues are therefore determined implicitly by
\begin{equation}
P_m(x)=0,
\end{equation}
with no closed-form expression analogous to Eq.~\eqref{eq:cheb_roots}.

Consequently, while the eigenvalues can still be characterized implicitly via the recurrence relation, an explicit analytic expression analogous to the odd-size case is generally not available for even \(N\).

\subsection{Spectrum of odd stochastic and even quantum systems in 1D SSH chain}
\label{sec:1Dsshsize}
We further compare the spectral overlap between stochastic systems of odd size and quantum systems of even size to examine the correspondence between the clusters. While the two spectra largely overlap, we observe a systematic shift: the upper stochastic cluster lies slightly below its quantum counterpart, whereas the lower stochastic cluster lies slightly above the corresponding quantum cluster. In these half-integer–cell lattices, the topologically emerging state is absent (Fig.~\ref{fig:sshodd}).

\section{Characteristic wavelength of the TES eigenvector in the 1D SSH chain}
\label{sec:sshtes}

In contrast to the Hatano-Nelson chain discussed above, the 1D SSH chain does not admit a simple sinusoidal representation of the eigenvector. Instead, the structure can be obtained through an inductive argument. The resulting expression reproduces the numerical eigenvectors and confirms that the observed behavior is not a finite-size effect.

The topologically emerging state (TES) eigenvector is as given in Eq.~\eqref{eq:TES_eigenvector},

The alternating factor $n$ introduces a $\pi$ phase shift under $n\to n+1$, corresponding to the wave number
\begin{equation}
q = \pi.
\end{equation}

The characteristic wavelength is therefore, overall,
\begin{equation}
\lambda_{\mathrm{TES}} = \frac{2\pi}{q} = 2.
\end{equation}

The ratio $c/(c-1)$ determines only the geometric envelope through the factor
\(
\left\lfloor n/2 \right\rfloor
\)
and does not affect the oscillation wavelength. 

Similar to the Hatano–Nelson chain, in the limit of extreme non-reciprocity the spectral clusters compress toward the steady state and their eigenvalues converge toward a single point. In this regime, the cluster eigenvectors whose eigenvalues overlap with that of the TES develop the same spatial localization profile as the TES eigenvector, leading to a complete overlap between these modes. The remaining cluster eigenvectors, although still distinct from the TES, likewise collapse toward a common localization pattern as their eigenvalues collapse into a point. Thus, extreme non-reciprocity produces both spectral clustering and a convergence of eigenvector localization profiles (Fig.~\ref{fig:sshvec}).

\section{2D Non--Hermitian SSH Lattice Hamiltonian}
\label{sec:2DSSHana}

The SSH Lattice quantum Hamiltonian is adopted from previous literatures \cite{tang2021topo,Nelson2024} as
\begin{equation}
\begin{aligned}
\mathcal{A}_{\mathrm{2D}}
&=\sum_{i=0}^{N_i-1}\sum_{j=0}^{N_j-1}\Big[
\gamma_{\mathrm{in}}\big(
|(i,j)_B\rangle\langle(i,j)_C|
+|(i,j)_A\rangle\langle(i,j)_B|\\
&+|(i,j)_D\rangle\langle(i,j)_A|
+|(i,j)_C\rangle\langle(i,j)_D|
\big)\\
&\qquad\qquad
+\gamma'_{\mathrm{in}}\big(
|(i,j)_C\rangle\langle(i,j)_B|
+|(i,j)_B\rangle\langle(i,j)_A|\\
&+|(i,j)_A\rangle\langle(i,j)_D|
+|(i,j)_D\rangle\langle(i,j)_C|
\big)\Big]\\
&+\sum_{i=0}^{N_i-2}\sum_{j=0}^{N_j-1}\Big[
\gamma_{\mathrm{ex}}\big(
|(i+1,j)_C\rangle\langle(i,j)_B|\\
&+|(i,j)_A\rangle\langle(i+1,j)_D|
\big)\\
&+\gamma'_{\mathrm{ex}}\big(
|(i,j)_B\rangle\langle(i+1,j)_C|
+|(i+1,j)_D\rangle\langle(i,j)_A|
\big)\Big]\\
&+\sum_{i=0}^{N_i-1}\sum_{j=0}^{N_j-2}\Big[
\gamma_{\mathrm{ex}}\big(
|(i,j+1)_B\rangle\langle(i,j)_A|\\
&+|(i,j)_D\rangle\langle(i,j+1)_C|
\big)\\
&+\gamma'_{\mathrm{ex}}\big(
|(i,j)_A\rangle\langle(i,j+1)_B|
+|(i,j+1)_C\rangle\langle(i,j)_D|
\big)\Big],
\end{aligned}
\label{eq:A2D}
\end{equation}

where $i=1,\ldots,N_i$ and $j=1,\ldots,N_j$.

We then use the relationship $\mathcal{W}=\mathcal{A}-\mathcal{D}$, where $\mathcal{D}_{ij}=\delta_{ij}\sum_k\mathcal{A}_{ki}$ \cite{schnakenberg1976} to derive the stochastic matrix.

\bibliographystyle{apsrev4-2}
\bibliography{references}
\end{document}